\documentclass[twocolumn,prd,noshowpacs,nofootinbib,amsmath,amssymb,superscriptaddress]{revtex4}
\usepackage{graphicx,epsfig,psfrag,bm,amssymb}
\usepackage{dcolumn}
\usepackage{bm}
\usepackage{mciteplus}
\usepackage{color}
\usepackage{mathrsfs,amsmath, amsfonts,hepunits, color}
\usepackage{mciteplus}
\usepackage{tikz}
\usepackage{slashed}
\usetikzlibrary{arrows,shapes}
\usetikzlibrary{trees}
\usetikzlibrary{matrix,arrows} 				% For commutative diagram
\usepackage{slashed}

\newcommand{\be}{\begin{eqnarray}}
\newcommand{\ee}{\end{eqnarray}}
\def\lsim{\mathrel{\rlap{\lower4pt\hbox{\hskip 0.5 pt$\sim$}}
    \raise1pt\hbox{$<$}}}                % less than or approx. symbol
\def\gsim{\mathrel{\rlap{\lower4pt\hbox{\hskip1pt$\sim$}}
    \raise1pt\hbox{$>$}}}

\def\lsim{\mathrel{\rlap{\lower4pt\hbox{\hskip1pt$\sim$}}
    \raise1pt\hbox{$<$}}}
\def\gsim{\mathrel{\rlap{\lower4pt\hbox{\hskip1pt$\sim$}}
    \raise1pt\hbox{$>$}}}

\newcommand{\Zp}{{\rm Z'}}
\newcommand{\ZZ}{{\rm Z^0}}
\newcommand{\WW}{{\rm W^\pm}}

\newcommand{\HH}{{\rm H}}

\newcommand{\massdrop}{\frac{m_1}{m_{12}}}
\newcommand{\scaleconst}{\zeta_c}
\newcommand{\observable}{\zeta}

\begin{document}

\title{Improving Identification of Dijet Resonances at Hadron Colliders}
\author{Eder Izaguirre}
 \affiliation{Perimeter Institute for Theoretical Physics, Waterloo, Ontario, Canada}
\author{Brian Shuve}
 \affiliation{Perimeter Institute for Theoretical Physics, Waterloo, Ontario, Canada}
  \affiliation{Department of Physics, McMaster University,  
Hamilton, ON, Canada}
\author{Itay Yavin}
 \affiliation{Perimeter Institute for Theoretical Physics, Waterloo, Ontario, Canada}
 \affiliation{Department of Physics, McMaster University,  
Hamilton, ON, Canada}

\begin{abstract}
The experimental detection of resonances has played a vital role in the development of subatomic physics. The overwhelming multi-jet backgrounds at the Large Hadron Collider (LHC) necessitate the invention of new techniques to identify resonances decaying into a pair of partons. In this $\textit{Letter}$ we introduce an observable that achieves a significant improvement in several key measurements at the LHC: the Higgs boson decay to a pair of b-quarks; $\WW/\ZZ$ vector-boson hadronic decay; and extensions of the SM with a new hadronic resonance. Measuring the Higgs decay to b-quarks is a central test of the fermion mass generation mechanism in the SM, whereas the $\WW/\ZZ$ production rates are important observables of the electroweak sector.  Our technique is effective in large parts of phase-space where the resonance is mildly boosted and is particularly well-suited for experimental searches  dominated by systematic uncertainties, which is true of many analyses in the high-luminosity running of the LHC. 
\end{abstract}

\maketitle

%%%%%%%%%%%%%%%%
% Introduction
%%%%%%%%%%%%%%%%
{\noindent \it \bf Introduction \---}
Hadron colliders open a direct observational window on physics at the shortest distance scales. Indeed, the heaviest fundamental particles in the SM, the top quark and the Higgs boson, were discovered at the Tevatron and LHC, respectively \cite{Abachi:1995iq,Abe:1995hr, Aad:2012tfa, Chatrchyan:2012ufa}. However, hadronic collisions also give rise to large rates for jet-rich processes from Quantum Chromodynamics (QCD), which can mimic those of a non-QCD origin, such as jets from the decays of electroweak gauge and Higgs bosons. As an alternative, many of the properties of the SM bosons are typically measured in final states with one or more leptons, at the expense of smaller signal rates resulting from the subdominant leptonic branching ratios of the $\HH$, $\WW$, and $\ZZ$. In scenarios where there is a discrepancy between experimental results in fully leptonic channels and the SM predictions, or where the signal rate is very small, it is important to consider hadronic decays of electroweak bosons. Moreover, some theories beyond the SM (BSM) predict the existence of new particles that decay entirely hadronically, such as supersymmetric theories with hadronic $R$-parity violation  \cite{Weinberg:1981wj,Sakai:1981pk,Hall:1983id, Barbier:2004ez}.
Any improvement in the discrimination of hadronic resonances from QCD backgrounds further maximizes a hadron collider's ability to perform precision SM measurements and its  potential to discover new resonances.

There are several examples which motivate an improvement in the identification of hadronic resonances. The measurement of the production cross section of ${\rm WW+WZ}$ in the semi-leptonic channel can help shed light on the persistent discrepancy between theoretical predictions of the ${\rm WW}$ production cross section and experimental results from the fully leptonic channel~\cite{ATLAS:2012mec,Chatrchyan:2013yaa,Curtin:2013gta,Meade:2014fca,Jaiswal:2014yba}. Another example is the precision measurement of the Higgs boson decay rate into a pair of b-quarks, which is an essential test of the Higgs mechanism for generating fermion masses. Moreover, many theoretical extensions of the SM predict new resonances which decay into electroweak bosons and/or directly into hadrons. An improvement in the ratio of signal to background rates is important since searches for hadronic final states are often limited by systematic uncertainties, particularly with high integrated luminosity.

One region of phase space where the kinematics of the signal differs substantially from that of the background is the boosted resonance regime where the decay products are collimated. They can then be grouped into a single jet whose substructure is dramatically different from that of jets originating from QCD emission. Building on earlier work~\cite{Seymour:1993mx,Butterworth:2002tt}, this approach was successfully applied by Butterworth et al.~\cite{Butterworth:2008iy} for the important case of the Higgs boson decaying into a pair of $b$-quarks. Their method (BDRS) relies on the fact that the masses of the two leading subjets originating from the b-quarks  tend to be much smaller than the mass of the entire jet, which reconstructs the Higgs. This is in contrast with jets from QCD processes, where the mass of the jet arises from a sequence of wide-angle emissions. Therefore, the ``mass-drop'' between the total jet mass and the subjet masses provides a powerful discriminant between boosted Higgs bosons and QCD jets. Subsequently, variants on this method have been used to identify other boosted hadronically decaying objets, such as top quarks and $\WW$ bosons \cite{Kaplan:2008ie, Plehn:2009rk, Ellis:2009me,Thaler:2008ju, Thaler:2010tr,Soper:2011cr,Hook:2011cq,Larkoski:2014wba,Larkoski:2013eya,Jankowiak:2011qa}.

While jet substructure tools are powerful for identifying boosted resonances, these techniques typically involve paying a large penalty in signal efficiency, since many SM and BSM processes produce hadronic resonances without a substantial boost. With smaller boost, partons from the resonance decay are no longer collimated to the same degree and can only be grouped into a single jet if the radius parameter $R$ of sequential jet clustering algorithms \cite{Salam:2009jx} is very large, since $R\propto m_{\rm resonance}/p_{\rm T}$. There can be several issues with this, among them the fact that contamination from the underlying event and pile-up grows with radius \cite{ATLAS:2012am}, and that defining new physics objects associated with the large-$R$ jets requires a re-calibration of detector response and other experimental effects (although see \cite{Nachman:2014kla}).

In this {\it Letter} we propose an alternative approach to identifying hadronically decaying resonances which is not restricted to the highly boosted regime. It draws its efficacy by scaling the original ``mass-drop'' with the separation between the two resolved jets, thus maintaining discriminating power over a wider range of dijet angular separation. Our technique involves only separately resolved jets whose properties and detector response have been well-measured at the LHC, and does not require the calibration of new objects. We demonstrate the effectiveness of this new technique in identifying dijet resonances in a variety of processes, both in the SM as well as its extensions.  Earlier studies have also considered the possibility of combining resolved and boosted analyses~\cite{Gouzevitch:2013qca}.

%%%%%%%%%%%%%%%%
% New observable
%%%%%%%%%%%%%%%%
{\noindent \it \bf New Observable \---}
Consider  the decay of a mildly boosted boson (${\rm V}$) into quarks, which are consequently well separated, and result in two resolved jets, $j_{1}$ and $j_{2}$.  We can use the resolved jets to form the standard ``mass-drop" variable, $\massdrop \equiv {\rm max}(m_{j_1},m_{j_2})/m_{12}$ where $m_{j_i}$ is the mass of $j_i$ and $m_{12}$ is the invariant mass of the dijet system. This is a useful variable in jet substructure studies, but for well-separated jets, a major component of the background comes from a hard QCD splitting, giving rise to a mass drop that decreases as the separation increases.
We therefore modify the conventional mass-drop criterion by considering cuts which become stricter as the separation of the jets increases, specifically imposing upper cuts on $m_1/m_{12}$ that scale with $1/\Delta R_{12}$.

We now give a heuristic argument for this form of cut by considering the relative ÒtypicalÓ scaling of signal and background. The mass of each individual jet arises from QCD radiation with average  $\langle m_j^2 \rangle = C \alpha_s R^2 p_T^2/\pi $ in the small jet radius $R$ limit through $\mathcal{O}(\alpha_s)$, where $p_T$ is the jet's transverse momentum and $C$ is a color-dependent and jet-algorithm-dependent order-unity factor~\cite{Ellis:2007ib}. Since $R$ is fixed, and for mild boosts $p_T\sim m_{\rm V}/2$, the individual jet mass for signal varies over only a limited range. On the other hand, the combined mass, $m_{12}$ should reconstruct the decaying ${\rm V}$ mass. By contrast, the background scaling can be very different. For example, consider $\rm{W}$+jets processes, where the extra jets can arise from a hard splitting from a single parton. Since the dijets arise from a hard splitting, we expect the combined mass to scale as $\langle m_{12}^2 \rangle = C \alpha_s \Delta R_{12}^2 P_T^2/\pi$, where $\Delta R_{12}$ is the separation between the individual jets, and $P_T$ is the transverse momentum of the combined system. This relation will of course break down for very large $\Delta R_{12}$, but we expect it to still be useful in the intermediate region between a boosted dijet system ($\Delta R_{12} < 0.4$) and one where the two jets are back-to-back ($\Delta R_{12} \sim \pi$).  We thus expect the following approximate scaling for the mass-drop variable,
\be
\massdrop &\propto& {\sqrt{\frac{C\alpha_s}{\pi}}}\, R \quad \quad {\rm signal} \\
\massdrop &\propto& \frac{R}{\Delta R_{12}} \quad \quad {\rm background} \label{eq:scaleddrop}
\ee
where $R$ is the jet radius used in defining the jets, often $R\approx 0.4$. 
The above relations motivate the following observable,
\be
\label{eqn:observable}
\observable \equiv \massdrop \times \Delta R_{12},
\ee
and the following cut as a discriminant of a dijet resonance against background, 
\be
\label{eqn:tagger_def}
\observable < \scaleconst,
\ee
with $\scaleconst$ a parameter optimized in Monte-Carlo studies. In what follows, we apply the cut, Eq.~(\ref{eqn:tagger_def}) to a variety of processes and find a consistent enhancement in signal over background ${\rm S/B} \sim {\rm 2-5}$. It comes at a price of a signal efficiency of $\mathcal{O}(10-20\%)$ and a small reduction in ${\rm S/\sqrt{B}}$. As such, it proves a powerful new tool to search for hadronic resonances in situations where the measurement is dominated by systematic uncertainties. We find that this observable significantly outperforms $\frac{m_1}{m_{12}}$ and $\Delta R_{12}$ individually (see Sec.~\ref{app:formulas} for comparison to other observables and demonstration of its robustness against MC and detector effects).

We note an alternative way of motivating the scaled mass-drop cut:~if one holds $m_{12}$ fixed, the maximum virtuality $Q_{\rm max}^2$ of the signal partons is fixed to be $(m_{12}/2)^2$, as can be seen in the rest frame of the resonance, while the background favours asymmetric splittings so that $Q_{\rm max}^2\sim m_{12}^2$. For fixed $m_{12}$, asymmetric splittings only dominate for small $\Delta R_{12}$, while at larger $\Delta R_{12}$ the jet $p_{\rm T}$ cuts make the background look signal-like with $Q_{\rm max}^2\sim (m_{12}/2)^2$, which is why a scaled mass-drop cut is effective.

While we have motivated a scaled mass-drop cut, the scaling with $\Delta R_{12}$ in Eq.~(\ref{eq:scaleddrop}) is only approximate. 
We study a generalization of the form,
\be
\label{eqn:gen_observable}
\observable(R_c) \equiv \massdrop \times \left(\Delta R_{12}- R_c\right)
\ee
with a similar cut as in Eq.~(\ref{eqn:tagger_def}). Here $R_c$ is again some phenomenological parameter optimized in Monte-Carlo. This generalized cut retains a higher signal efficiency at the price of only a small reduction in ${\rm S/B}$ relative to Eq.~(\ref{eqn:observable}).

Our variable depends on the small mass of a resolved R=0.4 jet which could be sensitive to pile-up contamination. We have checked that pile-up indeed affects the performance of our observable significantly; however we have verified that trimming~\cite{Krohn:2009th} with parameters optimized for small-radius jets  can recover the original sensitivity to within 10-20\%.

%%%%%%%%%%%%%%% FIGURE %%%%%%%%%%%%%%%%%%%%%
\begin{figure}[h]
\centering
\includegraphics[scale=0.5]{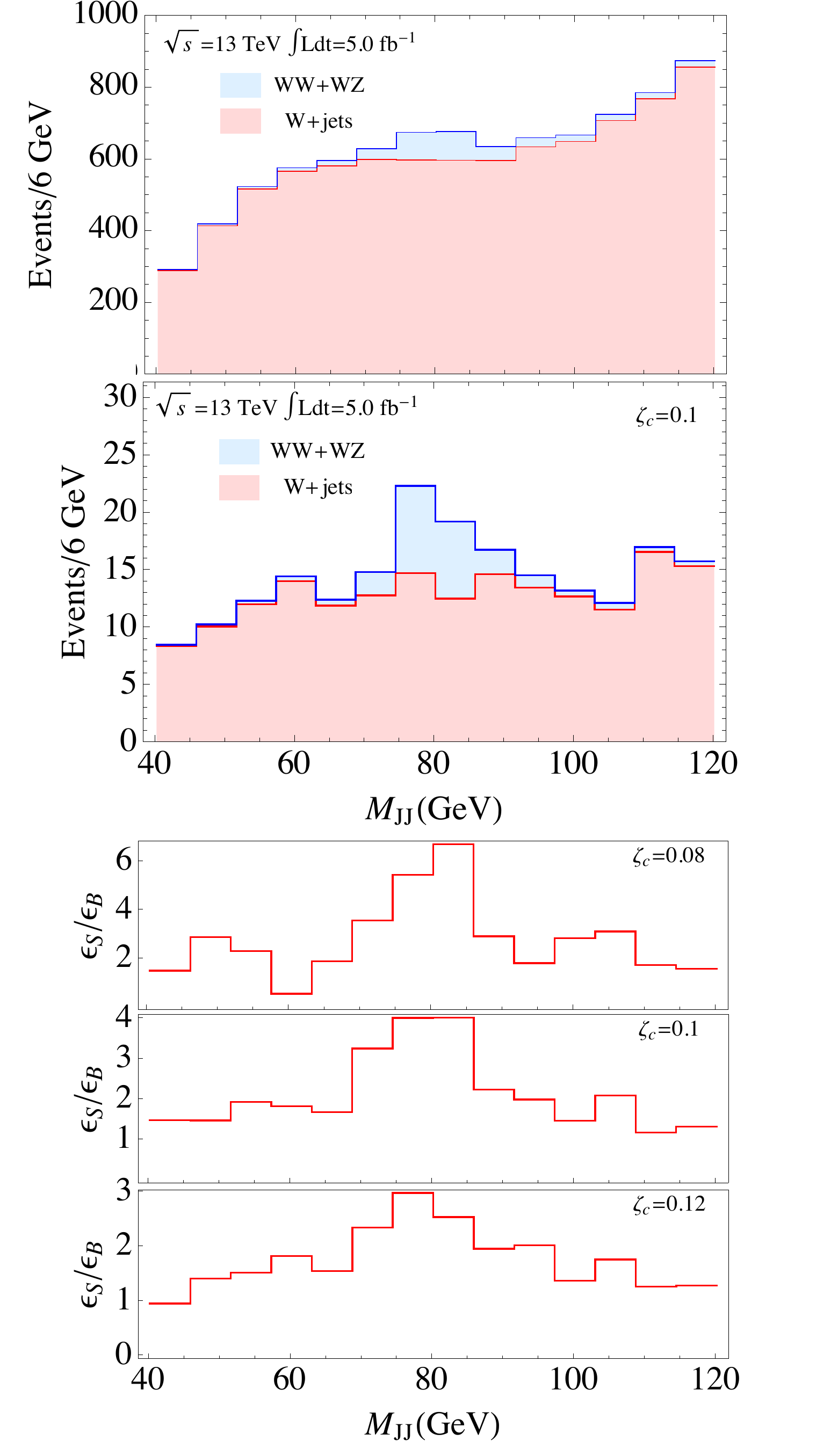}
\caption{The top pane shows histograms of both the dominant W+jets background as well as signal before applying the cut on $\observable$, but after all of the selection criteria described in the text. In the middle pane, we show the same distributions but after applying the $\observable$ cut. The bottom pane shows the binned ${\rm S/B}$ improvement for $\scaleconst = 0.08, 0.1,$ and $0.12$. The corresponding signal efficiencies are approximately $3\%, 10\%$, and $25\%$, respectively. }
\label{fig:WW_WZ}
\end{figure}
%%%%%%%%%%%%%%% FIGURE %%%%%%%%%%%%%%%%%%%%%

%%%%%%%%%%%%%%%%
% WW+WZ
%%%%%%%%%%%%%%%%
\vspace{2mm}
{\noindent \it \bf  WW+WZ \---}
As our first example, we consider the  measurement of the combined SM production cross section of $\rm WW+WZ$ in the semi-leptonic channel, i.e.~in $pp \rightarrow jj\ell \nu$, with the two jets reconstructing the $\WW$ or the $\ZZ$. This measurement is of particular importance given the persistent discrepancy between theory and observation in the $\rm WW$ cross section in the fully leptonic channel $pp\rightarrow \ell\nu\ell'\nu'$ \cite{ATLAS:2012mec,Chatrchyan:2013yaa,Curtin:2013gta,Meade:2014fca,Jaiswal:2014yba}. 

We follow the $7~\TeV$ analysis presented by the CMS collaboration in ref.~\cite{Chatrchyan:2012bd}, where by far the largest  background is $\rm W$+jets. We simulated both background and signal events using M\textsc{adgraph} 5~\cite{Alwall:2011uj} and showering through P\textsc{ythia} 6~\cite{Sjostrand:2006za}, with matching of matrix elements to parton showers done using an MLM-based shower scheme~\cite{Alwall:2008qv}. Jets were then clustered using the anti-$k_T$ algorithm~\cite{Cacciari:2008gp} with the F\textsc{astjet} package~\cite{Cacciari:2011ma}, and our simulations are in good agreement with the results of ref.~\cite{Chatrchyan:2012bd}. At $\sqrt{s}=13~\TeV$, we apply similar selection cuts as the CMS analysis but with the jet $p_T$ and $\slashed{E}_T$ cuts scaled up to $50~\GeV$. To quantify the gain obtained from an additional cut on $\observable$,  in Fig.~\ref{fig:WW_WZ} we plot the dijet invariant mass distribution before and after the $\observable$ cut. We also show the improvement in ${\rm S/B}$ in each dijet invariant mass bin. A gain in ${\rm S/B}$ of 3-4 is obtained with a cut of $\observable<0.1$ at the price of $\mathcal{O}(10\%)$ signal efficiency. This could improve the precision with which the ${\rm WW+WZ}$ cross section is measured at Run-II of the LHC at high luminosity.

%%%%%%%%%%%%%%%%
% VH(bb)
%%%%%%%%%%%%%%%%
\vspace{2mm}
{\noindent \it \bf Higgs decay to $b\bar{b}$ \---}
The second example we consider is the very important SM process $pp\rightarrow {\rm H}\WW\rightarrow b\bar{b}\ell\nu$. This
%%%%%%%%%%%%%%% FIGURE %%%%%%%%%%%%%%%%%%%%%
\begin{figure}[h]
\centering
%\includegraphics[scale=0.7]{figs/WH>bb_before_cut.pdf}\\
%\includegraphics[scale=0.7]{figs/WH>bb_after_cut.pdf}\\
%\vspace{2mm}
%\includegraphics[scale=0.68]{figs/WH>bb_SvsRc.pdf}\\
%\includegraphics[scale=0.68]{figs/WH>bb_SoBvsRc.pdf}
\includegraphics[scale=0.68]{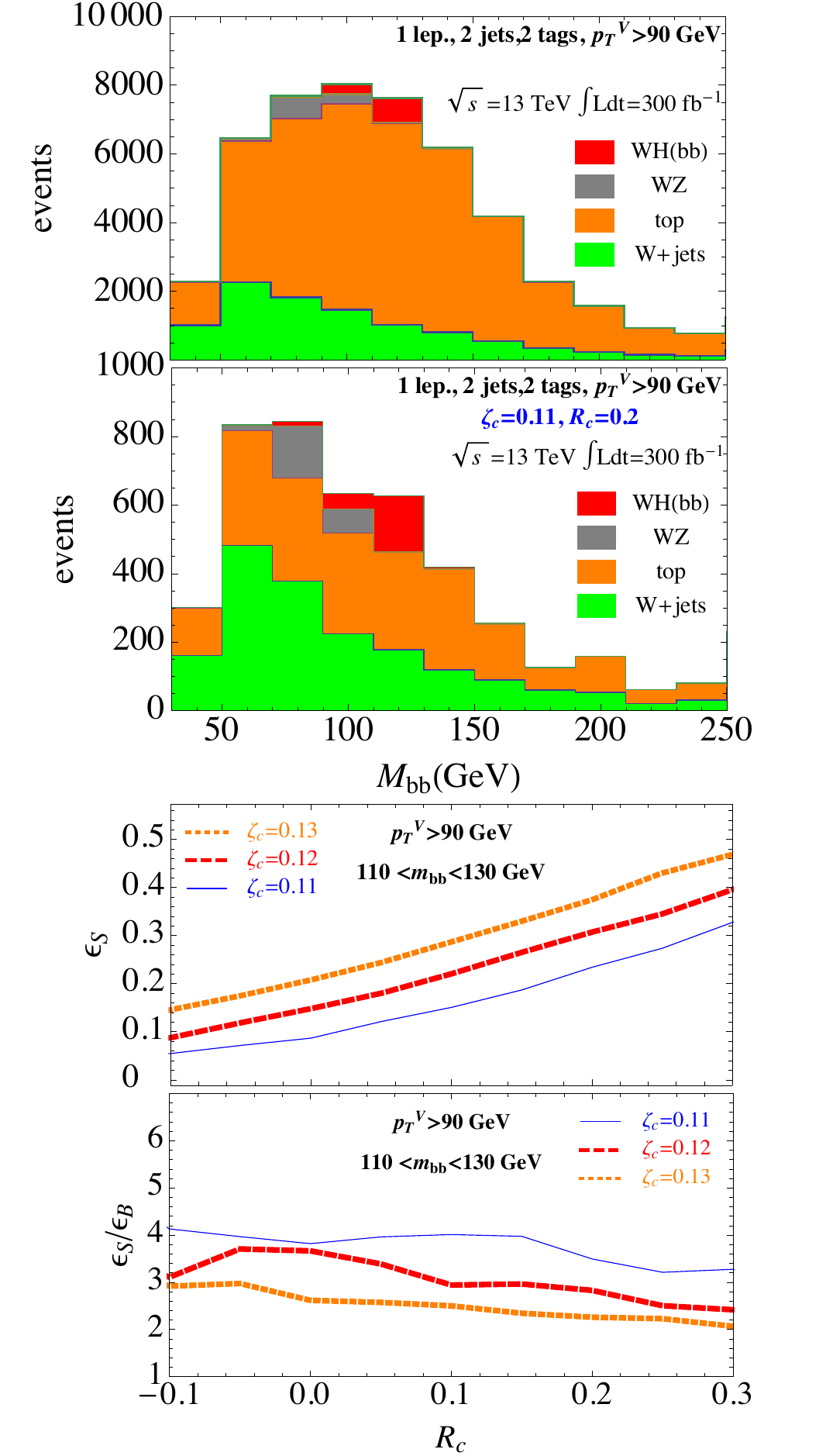}
\caption{In the upper two panes, we show the invariant mass of the $b$-tagged jets before and after the cut, Eq.~(\ref{eqn:gen_observable}),  for both signal and background. In the lower two panes, we show the relative signal efficiency as well as its ratio with background efficiency  as a function of $R_c$ in the mass window and $p_{\rm T}$ region indicated.}
\label{fig:WHtobb}
\end{figure}
%%%%%%%%%%%%%%% FIGURE %%%%%%%%%%%%%%%%%%%%%
 is a challenging rate to measure at the LHC and was the subject of the boosted analysis in ref.~\cite{Butterworth:2008iy}, which
 inspired much subsequent work into jet substructure.  We use it to demonstrate the utility of the shifted observable, 
Eq.~(\ref{eqn:gen_observable}), in maintaining a higher signal efficiency with a similar enhancement in S/B.

We follow the existing analysis by ATLAS~\cite{TheATLAScollaboration:2013lia} done at $7$ and $8~\TeV$ and validate our simulations by applying the same event selection criteria. In what follows, we concentrate on the higher energy run at $13~\TeV$ and apply the same selection. Both background and signal were simulated as discussed above. For concreteness, we focus on the WH signal with two resolved b-tagged jets and one lepton as outlined in the ATLAS analysis. We concentrate on the part of phase-space where the vector-boson transverse momentum is $p_{\rm _T}^{\rm _V}>90~\GeV$, and the Higgs boson consequently has a modest boost. The dominant backgrounds after $b$-tagging are top pair, $\rm{W}$+jets, and diboson production. Fig.~\ref{fig:WHtobb} displays the $b$-tagged dijet invariant mass distributions for both background and signal at $\sqrt{s} = 13~\TeV$ before and after applying the cut from Eq.~(\ref{eqn:gen_observable}) with $\scaleconst = 0.11$ and $R_c=0.2$. The parameter $\scaleconst$ is consistent with the examples above, whereas the parameter $R_c$ was chosen to yield an improvement in relative signal efficiency ($\sim 20\%$) while maintaining a similar increase in S/B of $3-4$. The lower panes in Fig.~\ref{fig:WHtobb} show the behavior of the relative signal efficiency and the efficiency ratio against the parameter $R_c$ for various choices of $\scaleconst$. One generally obtains a larger efficiency with increased $R_c$ and $\scaleconst$, at the price of a reduced gain in S/B. Importantly, it is possible to obtain an enhancement in S/B without sacrificing $\rm S/\sqrt{B}$. Even excluding the highly boosted region ($p_{\rm _T}^{\rm _V}>200~\GeV$), the cut on $\observable$ yields an enhancement of ${\rm S/B}$ of $2-3$. In that respect, our method yields gains in regions of phase-space complementary to that of BDRS, which targets highly boosted Higgs bosons. 

 Searches which are limited by statistical uncertainties are difficult to improve upon with our method since ${\rm S/\sqrt{B}}$ is typically not enhanced. The ATLAS $\mathrm{H}\rightarrow b\bar{b}$ analysis~\cite{TheATLAScollaboration:2013lia} which combines the full $7$ and $8~\TeV$ dataset has only about $\sim 30$ events in the region considered (two resolved b-tagged jets, one lepton, and $p_{\rm _T}^{\rm _V}>90~\GeV$). However, the goal of such a search is to determine the $\mathrm{H}b\bar{b}$ coupling as precisely as possible, and with the accumulation of more data, statistics will be less of a concern and our method may prove useful once systematic uncertainties dominate.

%%%%%%%%%%%%%%%% FIGURE %%%%%%%%%%%%%%%%%%%%%
%\begin{figure}[t]
%\centering
%%\includegraphics[scale=0.75]{figs/zp>WW_before_cut_7TeV.pdf}\\
%%\includegraphics[scale=0.75]{figs/zp>WW_after_cut_7TeV.pdf}\\
%%\includegraphics[scale=0.7]{figs/ZpToWW_analysis_LHC7_Seff.pdf}\\
%%\includegraphics[scale=0.7]{figs/ZpToWW_analysis_LHC7_SoB.pdf}
%\includegraphics[scale=0.76]{fig3.pdf}
%\caption{In the upper two panes we depict the invariant mass distribution of the $\mu\nu jj$ system before and after the cut, Eq.~(\ref{eqn:tagger_def}), for both signal and background. In the two lower panes we show the efficiency of the cut  and the ratio of signal to background efficiencies as a function of $\scaleconst$, for various values of the $Z'$ mass.}
%\label{fig:ZpToWW}
%\end{figure}
%%%%%%%%%%%%%%%% FIGURE %%%%%%%%%%%%%%%%%%%%%

\begin{table}[htdp]
\begin{center}
\begin{tabular}{|c|c|c|c|}
\hline
& $M_{\Zp}=600\GeV$ & $M_{\Zp}=800\GeV$ & $M_{\Zp}=1000\GeV$ \\
\hline
$\scaleconst=0.09$ & 4.5 (0.12) & 4.9 (0.27) & 3.4 (0.16) \\
$\scaleconst=0.11$ & 2.9 (0.29) & 3.1 (0.40) & 2.2 (0.29) \\
$\scaleconst=0.13$ & 2.0 (0.48) & 2.0 (0.54) & 1.6 (0.42)  \\
\hline
\end{tabular}
\caption{The ratio of signal to background efficiency gain for different values of $\scaleconst$ and the $\Zp$ mass. In brackets are the signal efficiencies. }
\label{tbl:ZpToWW}
\end{center}
\label{default}
\end{table}%

%%%%%%%%%%%%%%%%
% Z' --> WW
%%%%%%%%%%%%%%%%
\vspace{2mm}
{\noindent \bf ${\rm Z'\rightarrow WW}$ and ${\rm W'\rightarrow ZW}$ searches \---}
The final example we consider is of BSM physics with a new heavy resonance decaying into two vector bosons which subsequently decay semi-leptonically, ${\rm \Zp \rightarrow WW} \rightarrow jj\ell\nu$ or ${\rm W'\rightarrow WZ} \rightarrow jj\ell\nu$. These two processes have similar kinematics and so we focus on the former for concreteness.  This example allows us to investigate the performance of the cut as the boost of the hadronic ${\rm W}$ varies with the $\Zp$ mass (the boost is $\sim M_{_{\Zp}}/2M_{\rm _W}$).

We follow the $7~\TeV$ ATLAS analysis of ref.~\cite{Aad:2013wxa} and simulate both background and signal events as described above, concentrating on the dominant $t\bar{t}$ and ${\rm W}+$jets backgrounds. We apply all cuts from the ATLAS diboson analysis.  In Tbl.~(\ref{tbl:ZpToWW}), we show the ratio of signal to background efficiency gain for various values of $\scaleconst$ and the $\Zp$ mass. The efficiency is calculated in a window of $M_{jj\ell\nu}$ within $10\%$ of $M_{\Zp}$. The improvement in sensitivity is significant as for example a mass-point ($M_{\Zp} = 800\GeV$) which was only marginally excluded due to large ($\sim30\%$) systematic uncertainties, can be thoroughly ruled-out.

{\noindent \it \bf Conclusions \---} In this {\it Letter}, we introduced a new observable that improves the identification of resonances that decay to two resolved jets. It is the product of the ``mass-drop" variable known from jet substructure studies and the dijet separation, $\Delta R_{12}$. Importantly,  it uses only small-radius jets as employed by ATLAS and CMS, and a cut on this observable can be applied in a straightforward way to existing analyses without additional calibrations needed for large-radius clustering. We illustrated the efficacy of this observable and its generalization in enhancing S/B in important SM as well as BSM processes. The cut easily lends itself to optimization associated with the interplay between S/B enhancement and signal efficiency. Thus, it can be used in searches with different ratios of systematic to statistical uncertainties. Many more processes can benefit from this observable and we leave to future study the elucidation of its utility.

{\em Acknowledgments.}
We thank David Miller, Max Swiatlowski, and Scott Thomas for helpful discussions. This work was made possible by the facilities of the Shared Hierarchical 
Academic Research Computing Network (SHARCNET) and Compute/Calcul Canada.
BS is supported in part by the Canadian Institute of Particle Physics. IY is grateful to the Mainz Institute for Theoretical Physics (MITP) for its hospitality and support during the completion of this work. This research was supported in part by Perimeter Institute for Theoretical Physics. Research at Perimeter Institute is supported by the Government of Canada through Industry Canada and by the Province of Ontario through the Ministry of Research and Innovation.
%\bibliographystyle{apsrevM}
%\bibliography{UndergroundAccel}

\appendix

\section{Robustness of the Observable}

\label{app:formulas}

\begin{figure}[h]
\centering
\includegraphics[scale=0.46]{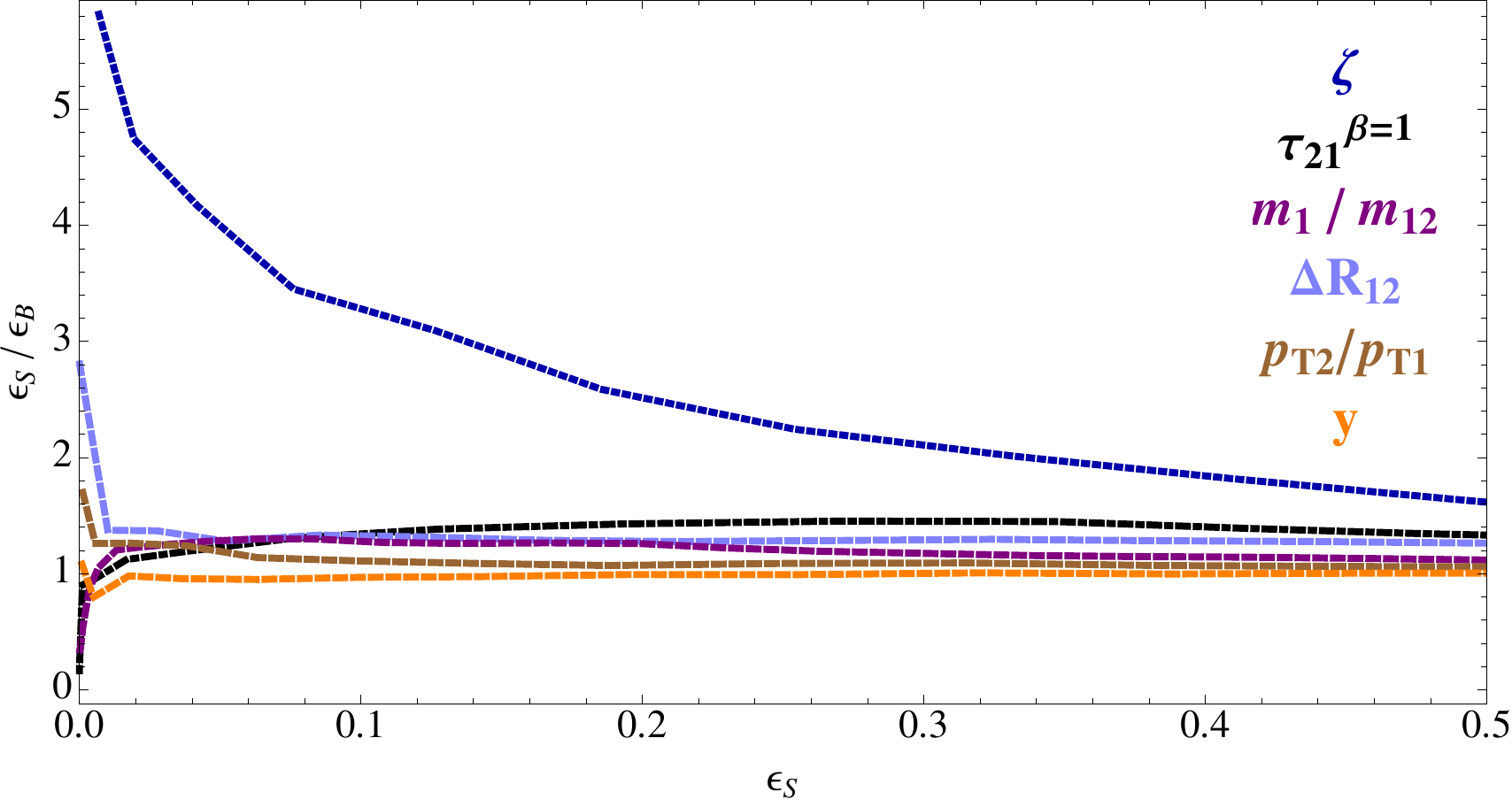}
\caption{A comparison of the gain in S/B relative to signal efficiency with cuts on different observables. }
\label{fig:cuts_comparison}
\end{figure}

In this section, we provide further evidence for the utility and robustness of the observable, Eq.~(3). The most immediate question is how this observable compares with other simple cuts on the dijet system or other observables known from jet-substructure studies. We focus on the WW+WZ search (using the same simulations and pre-selection cuts as in Fig.~1), but similar results hold for the other channels with some variation. In Fig.~\ref{fig:cuts_comparison} we plot the gain in S/B relative to the signal efficiency with the addition of a cut on the following observables derived from the two leading $R=0.4$ jets: $\observable=m_1 \Delta R_{12}/m_{12}$, defined in Eq.~(3); the $N$-subjettiness ratio $\tau_{21}^{\beta=1}$~\cite{Thaler:2010tr}, which is computed using a jet formed out of the constituents of the two leading jets; the mass drop, $m_1/m_{12}$; the angular separation, $\Delta R_{12}$; the ratio of the jet transverse momenta, $p_{\rm T2}/p_{\rm T1}$; and $y= p_{\rm T2}^2 \Delta R_{12}^2 / m_{12}^2$. The performance of $\observable$ is clearly superior to the other jet-substructure-inspired observables.

%%%%%%%%%%%%%%%%%
% Subsection
%%%%%%%%%%%%%%%%%

%%%%%%%%%%%%%%% FIGURE %%%%%%%%%%%%%%%%%%%%%
%%%%%%%%%%%%%%% FIGURE %%%%%%%%%%%%%%%%%%%%%

Since our observable depends on the masses of jets, which can be sensitive to the details of particular showering and hadronization models, another important question is how sensitive $\observable$ is to the choice of MC program and parameters. We again consider the WW+WZ search, generating parton-level events in \texttt{Madgraph 5} as described in the text, and we compare the performance of a cut on $\observable$ using events showered with \texttt{Pythia 6}, \texttt{Pythia 8}~\cite{Sjostrand:2007gs}, and \texttt{Herwig++}~\cite{Bahr:2008pv}. We use matrix element-parton shower matching using the shower-$k_\perp$ scheme for \texttt{Pythia 6}, the MLM scheme for \texttt{Pythia 8}, and we use an unmatched sample for \texttt{Herwig++}. The S/B gain from a cut on $\observable$ is comparable among the different MC generators, with the performance consistent to within $10-20\%$ for $\epsilon_S\gtrsim0.1$ considered in our paper. This demonstrates that $\observable$ is well-modelled by current MC programs, and is not subject to large theoretical uncertainties.

%%%%%%%%%%%%%%% FIGURE %%%%%%%%%%%%%%%%%%%%%
\begin{figure}[h]
\centering
\includegraphics[scale=0.6]{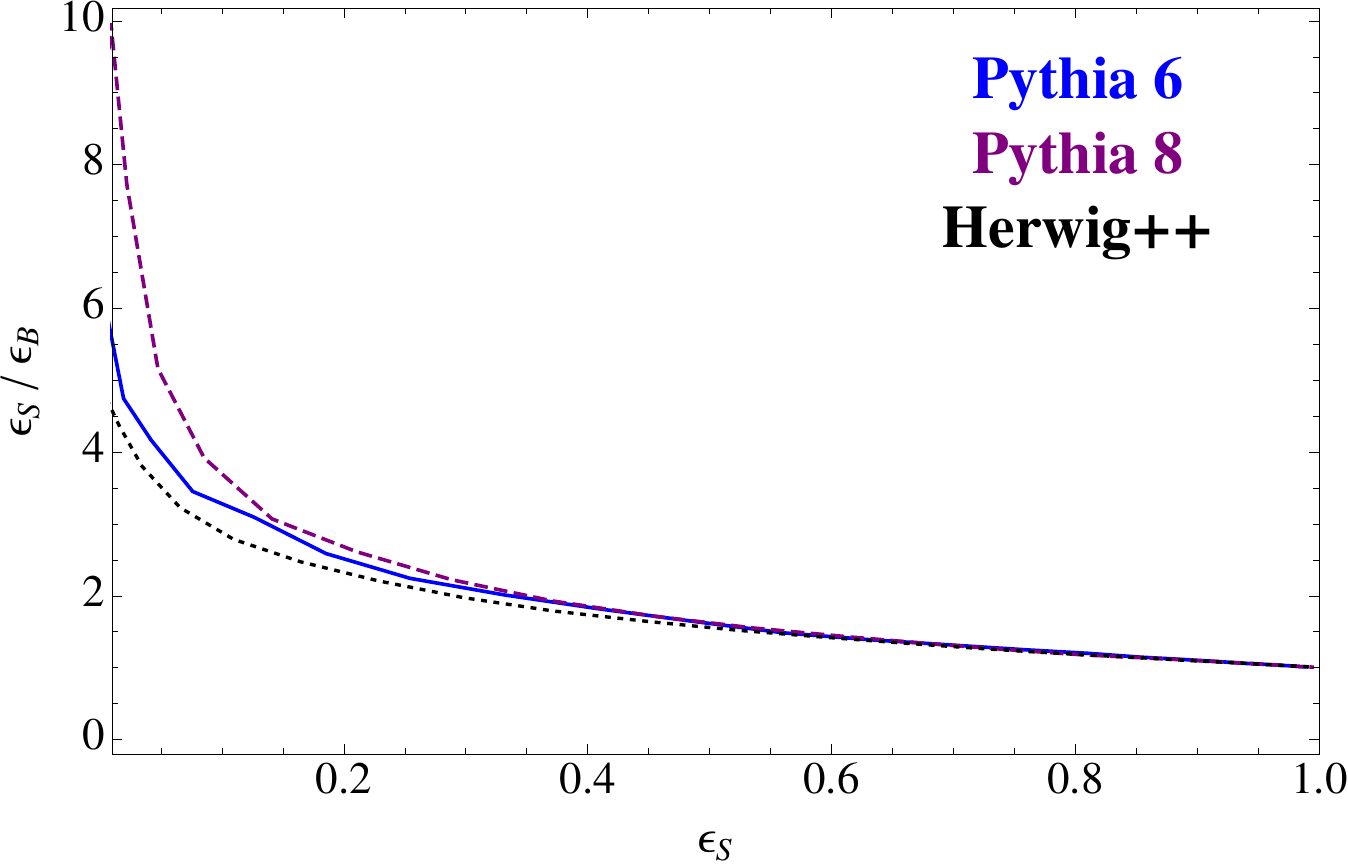}
\caption{A plot of the gain in S/B against the signal efficiency calculated with three different MC programs. A typical value for a cut on $\observable$ in our paper gives $\epsilon_S\approx0.1$. }
\label{fig:MC_comparison}
\end{figure}
%%%%%%%%%%%%%%% FIGURE %%%%%%%%%%%%%%%%%%%%%

\begin{figure}[h]
\centering
\includegraphics[scale=0.5]{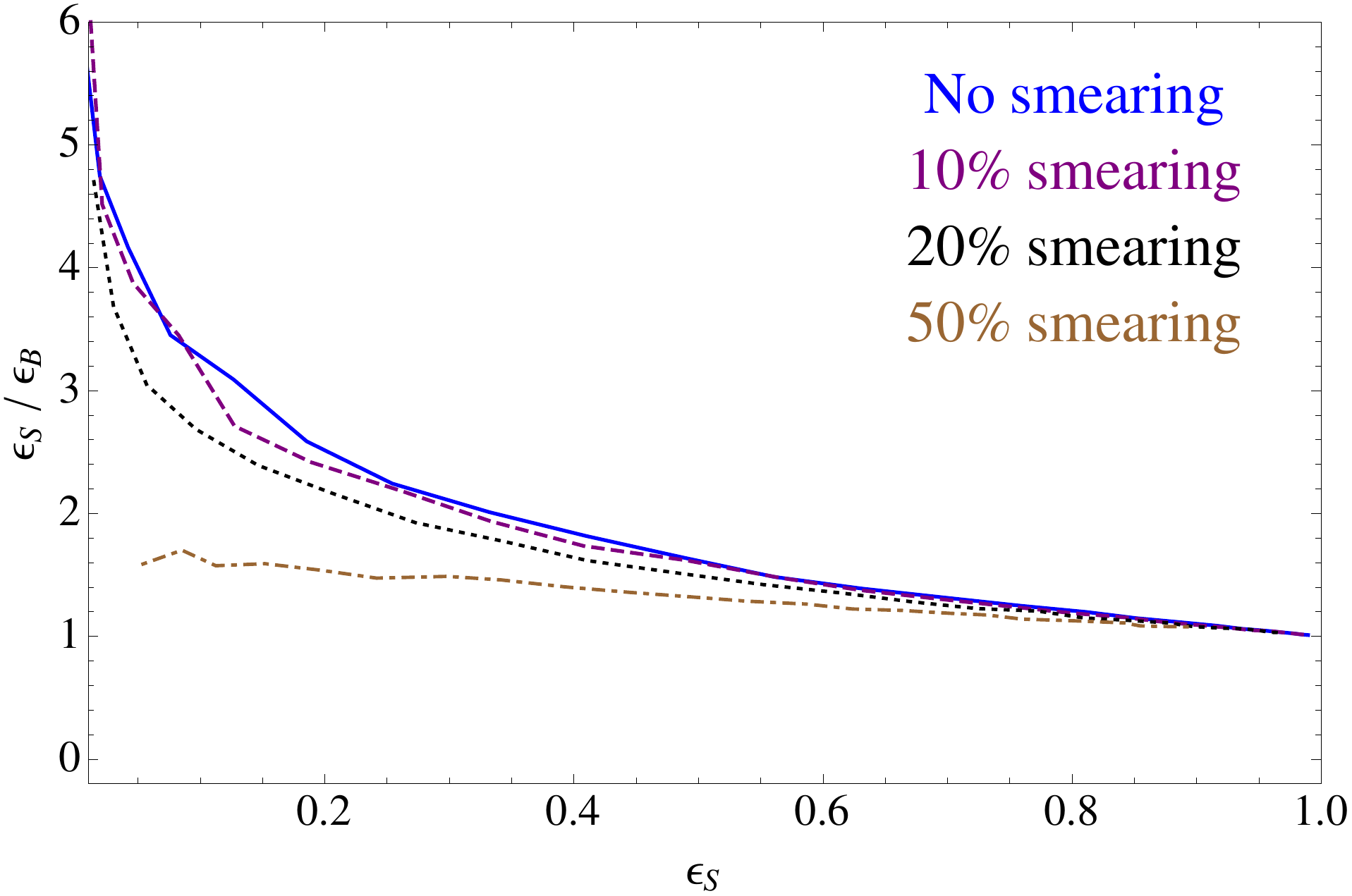}
\caption{A plot of the improvement of signal over background against the signal efficiency for various amounts of smearing applied to the resolved jet mass. Smearing is implemented as a simple random gaussian smearing with a width of $10, 20, $ and $50$\% of the reconstructed mass.}
\label{fig:smearing}
\end{figure}

In Fig.~\ref{fig:smearing} below we show the performance of the observable when different levels of smearing are applied to the the mass of the resolved jet. The separation between signal and background remains robust even after 20\% smearing of the jet mass. 

%%%%%%%%%%%%%%% FIGURE %%%%%%%%%%%%%%%%%%%%%
%%%%%%%%%%%%%%% FIGURE %%%%%%%%%%%%%%%%%%%%%

In the text, we presented an argument for why the background mass drop is larger than signal for fixed $m_{12}$. This is suggested by the fact that:
\begin{itemize}
\item For signal, the virtuality of the outgoing quarks is determined in the rest frame and is no more than $m_{12}/2$, half the resonance mass, which sets the scale for the typical final-state jet mass.
\item For background, the virtuality is determined by QCD splitting and can be as large as $m_{12}$, particularly for asymmetric splittings (small $\Delta R_{12}$). 
\end{itemize}

\begin{figure}[h]
\centering
\includegraphics[scale=0.78]{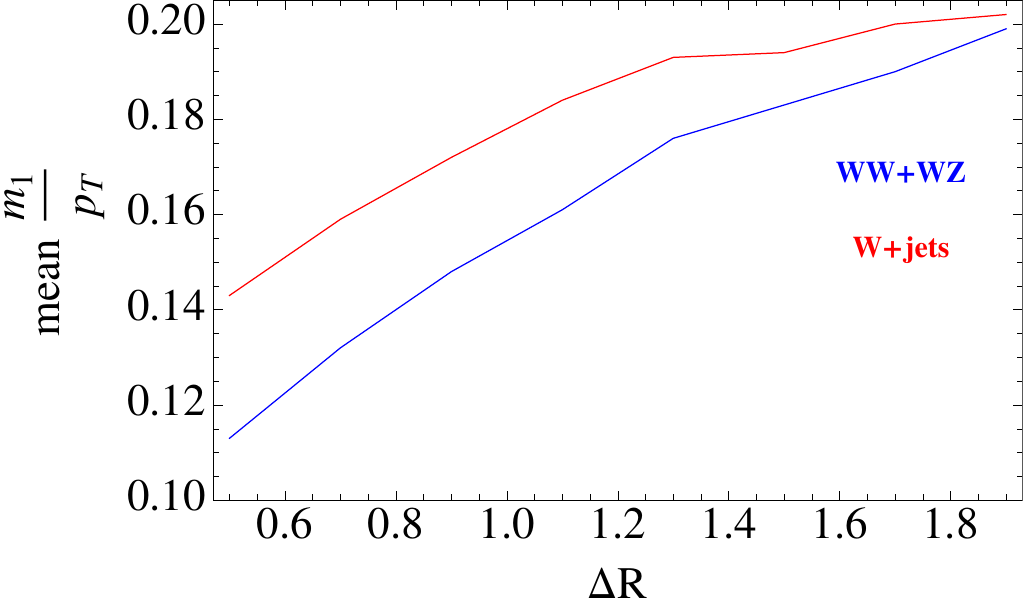}\hspace{1cm}\includegraphics[scale=0.75]{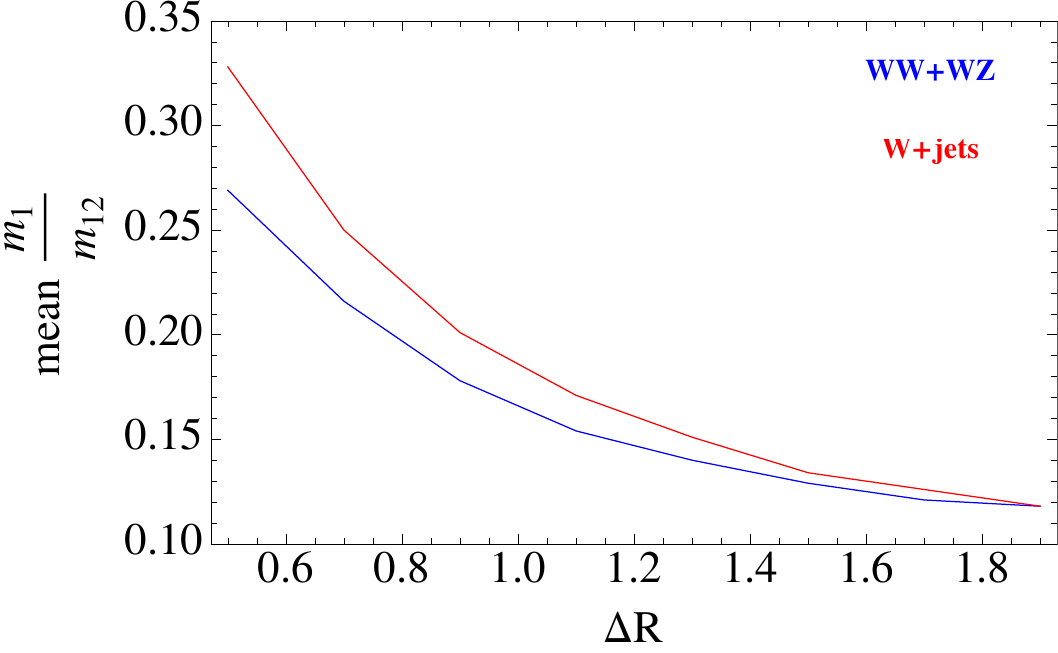}
\caption{A plot showing the average $m/p_{\rm T}$ (left pane) and $m/m_{12}$ (right pane) of the largest mass jet as a function of the distance $\Delta R_{12}$ between the two jets and fixed $m_{12}$. This  demonstrates that the typical mass of a signal jet is smaller than for a background jet when the kinematics are otherwise equal.}
\label{fig:m_pT}
\end{figure}

When fixing $m_{12}$ we generally expect $m_1/p_{T_1}$, the mass-to-$p_T$ ratio of the highest-mass jet, to be larger for background because that mass of that jet is determined by the higher virtuality. This is seen clearly in the left pane of Fig.~\ref{fig:m_pT}, where we show the differences between the average $m/p_{\rm T}$ between signal and background, which becomes more pronounced for smaller $\Delta R_{12}$ when the background splittings can become more asymmetric.  The result is that the mass drop is smaller for signal than background, particularly at small $\Delta R_{12}$ (see the right pane of Fig.~\ref{fig:m_pT}).

%%%%%%%%%%%%%%% FIGURE %%%%%%%%%%%%%%%%%%%%%
%%%%%%%%%%%%%%% FIGURE %%%%%%%%%%%%%%%%%%%%%

\bibliography{had_tagger_bib}

\end{document}